\documentclass[12pt,preprint]{aastex}
\usepackage{natbib}
\shorttitle{UV Spectrum of HD226868}
\shortauthors{Caballero-Nieves et al.}
\begin{document}


\title{The Ultraviolet Spectrum and Physical Properties of 
the Mass Donor Star in HD~226868 = Cygnus X-1\altaffilmark{1}}

\altaffiltext{1}{Based on observations with the NASA/ESA Hubble 
Space Telescope obtained at the Space Telescope Science Institute,
which is operated by the Association of Universities for Research
in Astronomy, Incorporated, under NASA contract NAS5-26555.
These observations are associated with programs GO-9646 and GO-9840.}

\author{S. M. Caballero-Nieves\altaffilmark{2}, D. R. Gies\altaffilmark{2,3}, 
C. T. Bolton\altaffilmark{4}, P. Hadrava\altaffilmark{5}, 
A. Herrero\altaffilmark{6}, T. C. Hillwig\altaffilmark{3,7}, 
S. B. Howell\altaffilmark{8}, W. Huang\altaffilmark{3,9},
L. Kaper\altaffilmark{10}, P. Koubsk\'{y}\altaffilmark{5}, 
and M. V. McSwain\altaffilmark{11}}

\altaffiltext{2}{Center for High Angular Resolution Astronomy,
Department of Physics and Astronomy, 
Georgia State University, P. O. Box 4106, Atlanta, GA  30302-4106;
scaballero@chara.gsu.edu, gies@chara.gsu.edu}
 
\altaffiltext{3}{Visiting Astronomer, Kitt Peak National Observatory,
National Optical Astronomy Observatory, operated by the Association
of Universities for Research in Astronomy, Inc., under contract with
the National Science Foundation.}

\altaffiltext{4}{Department of Astronomy and Astrophysics, 
University of Toronto, 50 St.\ George Street, Room 101, 
Toronto, Ontario, Canada M5S 3H4; 
bolton@astro.utoronto.ca}

\altaffiltext{5}{Astronomical Institute, Academy of Sciences of the Czech Republic, 
Fri\v{c}ova 298, CZ-251 65 Ond\v{r}ejov, Czech Republic; had@sunstel.asu.cas.cz, 
koubsky@sunstel.asu.cas.cz}

\altaffiltext{6}{Instituto de Astrof\'{i}sica de Canarias, 38200, La Laguna, Tenerife, Spain; 
Departamento de Astrof\'{i}sica, Universidad de La Laguna, 
Avda. Astrof\'{i}sico Francisco S\'{a}nchez, s/n, 38071 La Laguna, Spain;
ahd@ll.iac.es}

\altaffiltext{7}{Department of Physics and Astronomy, Valparaiso University, 
Valparaiso, IN 46383; todd.hillwig@valpo.edu}

\altaffiltext{8}{National Optical Astronomy Observatory, 
P. O. Box 26732, 950 N. Cherry Ave., Tucson, AZ 85719; howell@noao.edu} 

\altaffiltext{9}{Astronomy Department, University of Washington, Box 351580,
Seattle, WA 98195-1580; hwenjin@astro.washington.edu}

\altaffiltext{10}{Astronomical Institute Anton Pannekoek, Universiteit van Amsterdam, 
Kruislaan 403, 1098-SJ Amsterdam, The Netherlands; lexk@science.uva.nl}

\altaffiltext{11}{Department of Physics, Lehigh University, 
16 Memorial Drive East, Bethlehem PA 18015; 
mcswain@lehigh.edu}

\slugcomment{ApJ in press}
\paperid{}


\begin{abstract}

We present an examination of high resolution, ultraviolet spectroscopy
from {\it Hubble Space Telescope} of the photospheric spectrum of the
O-supergiant in the massive X-ray binary HD~226868 = Cyg~X-1.  We
analyzed this and ground-based optical spectra to determine the
effective temperature and gravity of the O9.7~Iab supergiant.  Using
non-local thermodynamic equilibrium (non-LTE), line blanketed, plane
parallel models from the TLUSTY grid, we obtain $T_{\rm eff} =
28.0\pm2.5$~kK and $\log g \gtrsim 3.00 \pm 0.25$, both lower than in
previous studies.  The optical spectrum is best fit with models that
have enriched He and N abundances.  We fit the model spectral energy
distribution for this temperature and gravity to the UV, optical, and
IR fluxes to determine the angular size of and extinction towards the
binary.  The angular size then yields relations for the stellar radius
and luminosity as a function of distance.  By assuming that the
supergiant rotates synchronously with the orbit, we can use the radius
-- distance relation to find mass estimates for both the supergiant
and black hole as a function of the distance and the ratio of stellar
to Roche radius.  Fits of the orbital light curve yield an additional
constraint that limits the solutions in the mass plane.  Our results
indicate masses of $23^{+8}_{-6} M_{\odot}$ for the supergiant and
$11^{+5}_{-3} M_{\odot}$ for the black hole.

\end{abstract}



\setcounter{footnote}{11}

\section{Introduction}                              

The massive X-ray binary Cygnus X-1 = HD~226868 consists of an O9.7
Iab primary \citep{wal73} with a black hole (BH) companion. The
fundamental properties of this system have been the subject of many
studies, but they continue to be controversial.  For example,
\citet{sha07} determined a relationship between black hole mass and
observed X-ray properties in the low frequency, quasi-periodic
oscillation -- spectral index plane to derive a BH mass of $8.7 \pm
0.8 M_{\odot}$ for Cyg~X-1, a value at the low end of previous
estimates \citep{gie86a,abu05}.  On the other hand, \citet{zio05} used
temperature -- luminosity relations in conjunction with evolutionary
models to calculate the mass of the bright, mass donor star.  Then,
with the orbital mass function from \citet{gie03} and the method
outlined by \citet{pac74}, he estimated the mass of the BH as $13.5-29
M_{\odot}$, at the high end of prior estimates.

Our goal in this paper is to determine if mass estimates from these
two methods can be reconciled through a re-examination of the
supergiant's spectrum to determine the stellar temperature, mass, and
radius.  Shortly after the X-ray source Cyg~X-1 was identified with
the star HD~226868 \citep{bol72,web72}, \citet{wal73} classified it as
a normal O9.7~Iab star.  The stellar temperature of a star of this
type depends critically upon the model atmosphere assumptions adopted
to match the line spectrum \citep{mar05}.  From a classical curve of
growth analysis of the optical spectrum of HD~226868, \citet{can95}
estimated an effective temperature of $T_{\rm eff} = 32\pm 2$~kK, and
they found an overabundance of He in the photosphere. \citet{her95}
also estimated the temperature of the star $T_{\rm eff} \approx 32$~kK
based upon fits of the optical line spectrum with calculated profiles
from unified model atmospheres that included a non-LTE treatment of H
and He but neglected line-blanketing from transitions of heavier
elements.  In addition, they determined values for gravity from fits
of the Balmer lines that ranged from $\log g = 3.03$ for
plane-parallel models to $\log g = 3.21$ for spherical models that
included wind effects.  Their results led to mass estimates of $17.8$
and $10.1 M_{\odot}$ for the supergiant and BH, respectively.  More
recently, \citet{kar08} classified HD~226868 as an ON star with a
temperature of $T_{\rm eff} = 30.4 \pm 0.5$~kK and gravity of $\log g
= 3.31 \pm 0.07$ using a semi-gray model atmosphere that accounts for
non-LTE effects in some lines and for X-ray illumination.

Here we present an analysis of the photospheric parameters for the
supergiant based upon ground-based optical spectra and
high-resolution, UV spectra from the {\it Hubble Space Telescope}
Space Telescope Imaging Spectrograph ({\it STIS}).  These {\it STIS}
spectra were first presented by \citet{gie08} and \citet{vrt08} in
discussions of the orbital variations observed in the stellar wind
lines.  We compare the optical and UV line profiles of HD~226868 with
synthetic spectra based on line blanketed, non-LTE photospheric models
in order to determine the stellar temperature and gravity (\S2).
Since the continuum flux and spectral lines of the supergiant could be
influenced by X-ray heating, we search for heating effects in the
orbital UV flux variations using the low/hard state {\it International
Ultraviolet Explorer (IUE)} archival spectra and the high/soft state
{\it HST} spectra (\S3).  A stellar radius -- distance relation can be
determined from fits of the spectral energy distribution.  We use the
observed flux distribution and spectra of field stars in the same
region of the sky to estimate the reddening and extinction in the
direction of Cyg~X-1 and to determine the angular size of the star
(\S3).  Finally, we use this radius -- distance relation with the
method developed by \citet{pac74} to set mass limits as a function of
distance and to estimate the probable masses using constraints from
the rotational line broadening and ellipsoidal light curve (\S4).


\section{Ultraviolet and Optical Line Spectrum}     

We need to rely on the line spectral features to estimate temperature
since the UV and optical continuum falls in the long wavelength,
Rayleigh-Jeans part of the flux distribution where the shape of the
continuum is insensitive to temperature.  Some of the best line
diagnostics for late O-supergiants are found in the optical spectrum
where several ionization state line ratios and the Balmer line
profiles change dramatically with temperature and gravity
\citep{wal90,sea08}.  In this section, we use $\chi^2_{\nu}$ fits of
the optical and UV spectra with model spectra to estimate $T_{\rm
eff}$ and $\log g$.  Our data consist of high resolution UV spectra
taken with the {\it HST}/STIS (G140M grating, resolving power
$R=14500$) and two sets of optical spectra from the Kitt Peak National
Observatory Coud\'{e} Feed telescope (CF; 3759 -- 5086 \AA , $R=2990$)
and 4~m Mayall Telescope and RC spectrograph (RC; 4182 -- 4942 \AA ,
$R=5700$).  Details of these observations are given in Table~1 of
\citet{gie08}.  All of these flux-rectified spectra were shifted to
the rest frame (using the orbital solution given by \citealt{gie08})
and co-added to increase the signal-to-noise ratio.

We compared these spectra with model spectra from the TLUSTY/SYNSPEC
codes given in the grids OSTAR2002 \citep{lan03} and BSTAR2006
\citep{lan07}.  The model atmospheres are based upon a plane parallel
geometry, solar abundances, line blanketed opacities, and non-LTE
calculations of atomic populations for H, He, and representative atoms
up to Fe.  The model spectra are presented as a function of four
parameters: the microturbulent velocity of gas in the line forming
region, $\xi$, the stellar effective temperature, $T_{\rm eff}$,
logarithm of the gravitational acceleration in the photosphere, $\log
g$, and the chemical abundance of the gas.  The model spectra were
transformed to the observed wavelength grids by wavelength integration
and convolution with rotational and instrumental broadening functions.
We adopted a projected rotational velocity of $V\sin i = 98$
km~s$^{-1}$ \citep{gie86a}, linear limb darkening coefficients from
\citet{wad85}, and Gaussian representations of instrumental broadening
using the projected slit FWHM \citep{gie08}.

It was clear from inspection that a large microturbulence is required
to match the observed and model spectra.  The OSTAR2002 grid uses $\xi
= 10$ km~s$^{-1}$ throughout while the BSTAR2006 grid adopts $\xi = 2$
km~s$^{-1}$ for the full grid and $\xi = 10$ km~s$^{-1}$ for a
selection of low gravity (supergiant) models.  We found that the best
fits were obtained in all three spectral bands with the $\xi = 10$
km~s$^{-1}$ models, and this was especially true in the FUV where the
observed deep spectral lines were not matched with the lower
microturbulent velocity models.  An atmospheric microturbulence of
$\xi = 10$ km~s$^{-1}$ is typical for late-O supergiants \citep{rya02},
and \citet{can95} derived an estimate of $\xi = 10.7$ km~s$^{-1}$ from
a curve of growth analysis of the optical \ion{N}{3} lines in the
spectrum of HD~226868.

We tested the goodness-of-fit for each of the models of interest by
calculating the reduced $\chi^2_{\nu}$ statistic,
\begin{equation}\label{eq1}
 \chi^2_{\nu} = \sum^{N}_{i=0} \frac{[F_{obs}(\lambda_{i}) - F_{model}
 (\lambda_{i})]^2}{\sigma_{err}^2(\lambda_{i}) (N - 1)}.
\end{equation}
Here $N$ is the total number of wavelength points used in the fit and
$\sigma_{err} (\lambda_i)$ is the standard deviation of the mean
rectified flux (determined from the scatter at wavelength $\lambda_i$
among the individual spectra in the co-added mean).  We selectively
omitted from the summation spectral regions that contained stellar
wind features or interstellar lines that are not present in the model
spectra.  Our results are listed in Table~1 for a wide range of model
spectra with a microturbulence of $\xi = 10$ km~s$^{-1}$.  Column (1)
indicates the spectral region fit by ``HST'' for the FUV spectrum,
``CF'' for the KPNO CF, blue spectrum, and ``RC'' for the KPNO 4~m RC,
green spectrum.  Column (2) gives the grid value of gravity $\log g$,
and column (3) gives a code for the spectral model (``O'' for spectra
from the OSTAR2002 grid and ``B'' and ``BCN'' for spectra from the
BSTAR2006 grid).  Then follow 10 columns that list the measured
$\chi^2_{\nu}$ for grid values of $T_{\rm eff}$ (at increments of 2.5
kK and 1 kK for the OSTAR2002 and BSTAR2006 grids, respectively).

\placetable{tab1}      

The trends in Table~1 are represented in a combined contour diagram in
Figure~1.  Here the gray-scale contours represent the goodness-of fit
for the FUV spectrum, the solid lines for the blue spectrum, and the
dashed lines for the green spectrum.  Since there is not an exact
match between the predictions of the OSTAR2002 and BSTAR2006 grids
at their boundary, Figure~1 shows contours based only on the OSTAR2002
grid for high gravity models $\log g \geq 3.0$, while the contours in the
lower gravity region $\log g \leq 3.0$ are based upon the BSTAR2006 grid.
Note that there are no models available for high temperature, low
gravity region in the lower right part of the diagram because such
atmospheres approach or exceed the Eddington luminosity limit.  The
higher gravity, lower temperature region is empty because the
BSTAR2006 grid contains only models for a lower microturbulent
velocity $\xi = 2$ km~s$^{-1}$ for this parameter range.  Note that
the $\chi^2_{\nu}$ minima in Table~1 have values much larger than the
expected value of unity.  This is due to the inclusion of spectral
regions where there is evident mismatch because of incomplete removal
interstellar features, marginal differences in the continuum
placement, and real differences between the observed and models even
in the best fit cases.  For the purposes of intercomparison in
Figure~1, we have subtracted from each $\chi^2_{\nu}$ set the minimum
minus one value, so that the figure represents variance increases that
result only from changes in the assumed temperature and gravity.

\begin{figure}
\begin{center}
{\includegraphics[angle=90,height=12cm]{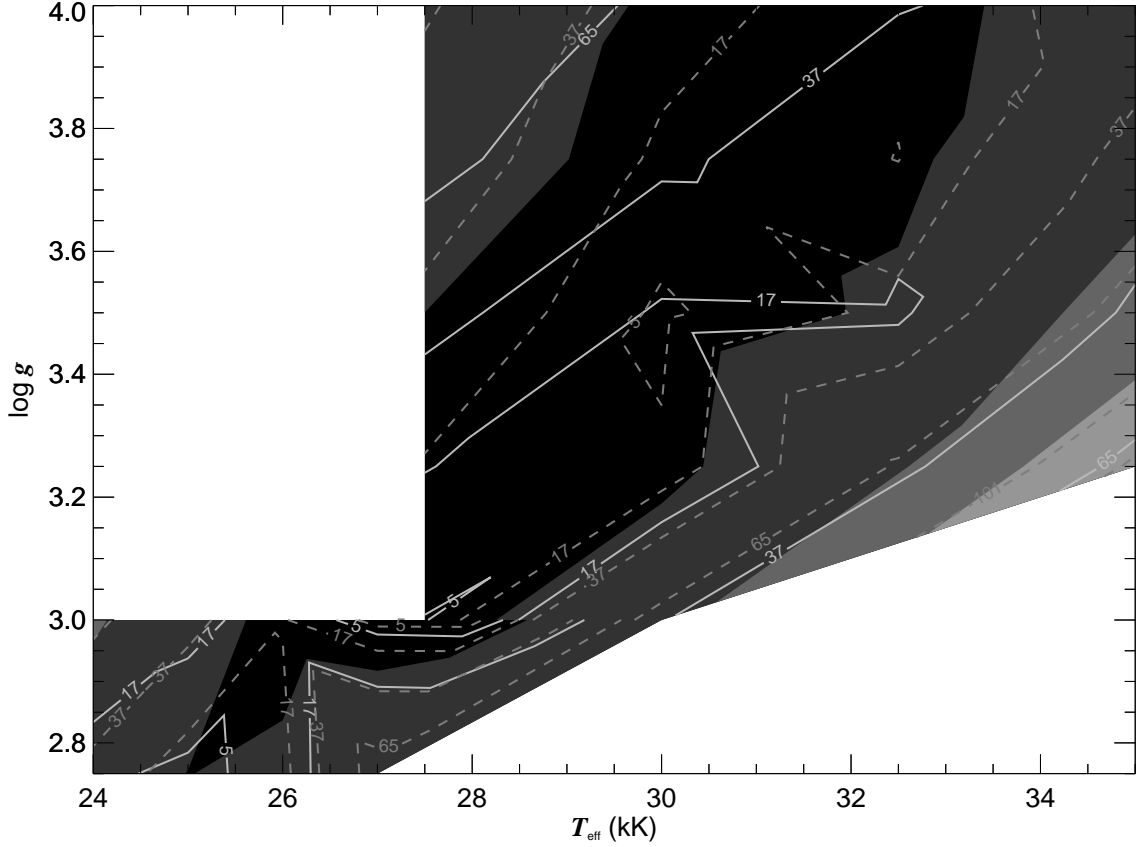}}
\end{center}
\caption{The variation within the $(T_{\rm eff}, \log g)$ plane of 
the net reduced $\chi^2_\nu$ statistic that measures the goodness of fit 
between the solar abundance models and the observed spectrum 
of HD~226868.  The contours show the value of $\chi^2_\nu$ above the 
best fit minimum (arbitrarily set to one), and they nominally 
represent the intervals of $2\sigma, 4\sigma, ... 10\sigma$ 
where $\sigma$ is the error in the parameter estimate. 
The contours for the FUV, KPNO CF, and KPNO 4 m spectral fits 
are shown as different gray-shaded regions, solid lines, and  
dashed lines, respectively.  The contours for $\log g \leq 3.0$ are
for fits with the BSTAR2006 models, while those for $\log g \geq 3.0$ are
for fits with the OSTAR2002 models.}
\label{fig1}
\end{figure}

\placefigure{fig1}     

The temperature and gravity properties of the $\chi^2_{\nu}$ fits
shown in Figure~1 result primarily from the dependence of the spectral
lines on the ionization levels in the gas.  The Saha ionization
equilibrium equation shows that the number ratio of atoms of one ion
to the next higher ionization state is equal to the electron density
times a function that decreases with increasing temperature.  Thus, in
order to match the ionization ratios represented in the spectral line
strengths, the best fits are found along a diagonal zone of increasing
$\log g$ with increasing temperature, i.e., increasing the electron
density with $\log g$ compensates for the functional drop related to
temperature increase.  However, there are other spectral dependences
that help us determine the minimum $\chi^2_{\nu}$ position along the
valley in the $(T_{\rm eff}, \log g)$ diagram.  In particular, the H
Balmer line wings in the optical spectrum are sensitive to pressure
broadening (linear Stark effect) and hence model gravity.  The lowest
contours of $\chi^2_{\nu}$ in Figure~1 indicate that the best fits are
found for $T_{\rm eff} = 26.5 - 28.5$ kK and $\log g = 2.9 - 3.1$.
Both of these estimates are lower than found in earlier studies
\citep{can95,her95,kar08}, but they are consistent with recent studies
that demonstrate that the inclusion of line blanketing in stellar
atmosphere models tends to lower the derived effective temperature
\citep{rep04,mar05,lef07,sea08}.

The run of $\chi^2_{\nu}$ fits shown in Figure~1 are for solar
abundance models, but given the reports that the spectrum of HD~226868
has strong N lines, we also explored fits with CN abundance models
included in the BSTAR2006 grid for gravities of $\log g = 2.75$ and
3.00.  These models assume a He to H number ratio of 0.2 (compared to
0.085 for the solar abundance models), a C abundance equal to one half
the solar value, and a N abundance five times the solar abundance.
These adjustments demonstrate the kind of changes that are expected
when the atmosphere becomes enriched in CNO-processed gas.  While the
CN models do not improve the fits of the FUV spectrum, they are
significantly better fits of the optical spectrum (where a number of
strong \ion{He}{1} and \ion{N}{3} lines are present; see Fig.~2
below).  When we compare the $\chi^2_{\nu}$ fit of a solar model to a
CN model at the same gravity in Table \ref{tab1}, the CN models fit
better at higher temperature, especially for fits of the optical
spectra. Following this trend, we estimate that the supergiant's
spectrum is fit best by the CN models with $T_{\rm eff}=28\pm2.5$ kK
and $\log g = 3.00\pm0.25$ dex. Thus, we will focus on these CN models
for the rest of this section.

We compare the two mean optical spectra (CF and RC) with our best fit
model spectrum in Figure~2 ($T_{\rm eff}=28$ kK, $\log g = 3.00$) and
with a marginally acceptable fit in Figure~3 ($T_{\rm eff}=26$ kK,
$\log g = 2.75$).  Also shown for comparison is the spectrum of a
similar O9.7~Iab star, $\mu$~Nor (HD~149038; from \citealt{wal90}).
The spectra appear in three plots: the short wavelength range from the
lower resolution CF data is shown in the top panel while the bottom
two panels illustrate the longer wavelength region from the higher
resolution RC spectra (compare with Fig.~1 in \citealt{kar08}).  In
general the agreement between the observed and model spectrum is
satisfactory.  The largest discrepancies are seen in the \ion{He}{2}
$\lambda 4686$ and H$\beta$ lines where incipient emission from the
stellar wind of the supergiant alters the profiles
\citep{gie86b,nin87a,gie03,gie08}.  The H Balmer line emission is
strongest in H$\alpha$, and for simple estimates of the Balmer
decrement for Case B recombination \citep{ost06}, we expect some
measurable degree of wind emission for all the Balmer lines shown in
Figures 2 and 3.  We find that the H line cores do appear shallower,
while the Balmer line wings agree well with the models.  \citet{sea08}
observed this effect in other supergiants and they suggest that
stellar wind emission from the outer atmosphere tends to fill in the
line core.  On the other hand, the Balmer line wings are formed in
higher density gas, deeper in the photosphere where we expect the
TLUSTY/SYNSPEC results to be quite reliable.  The H line wings become
narrower with lower gravity, and the predicted H profiles for the
lower gravity model illustrated in Figure~3 appear to be significantly
narrower than the observed ones.

\begin{figure}
\begin{center}
{\includegraphics[angle=90,height=12cm]{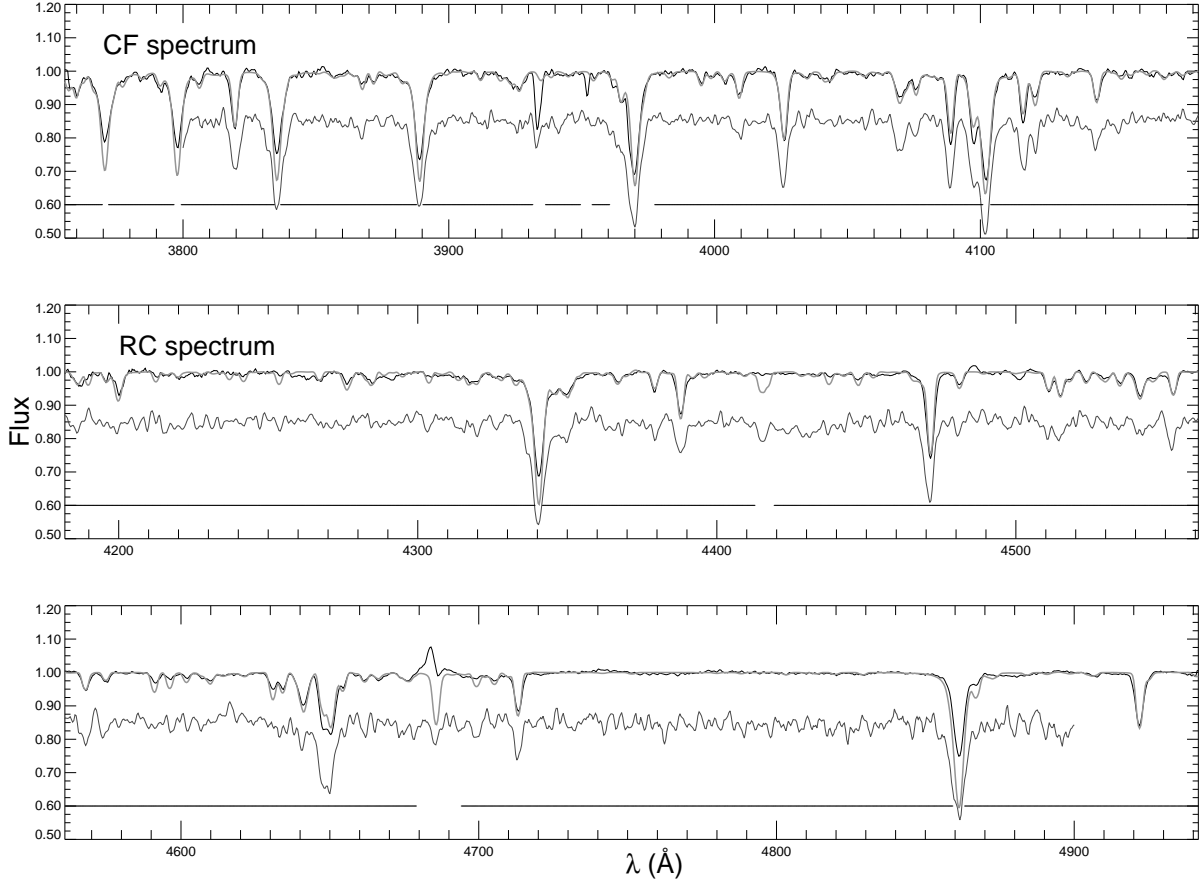}}
\end{center}
\caption{Rectified optical spectra ({\it dark line}) together with a
$T_{\rm eff} = 28$~kK and $\log g = 3.0$ TLUSTY CN model ({\it light
line}). The top panel shows the mean spectrum obtained with the KPNO
Coud\'{e} Feed telescope while the bottom two panels show the mean
spectrum obtained with the KPNO 4-m telescope.  The spectrum of the
O9.7~Iab star $\mu$~Nor is offset by 0.15 from the model and HD~226868
spectra for comparison. The horizontal lines below the spectra
indicate the wavelength regions included in the $\chi^2_\nu$
calculation. The \ion{He}{2} $\lambda4686$ emission line originates in
the focused wind from the star.}
\label{fig2}
\end{figure}

\placefigure{fig2}     

\begin{figure}
\begin{center}
{\includegraphics[angle=90,height=12cm]{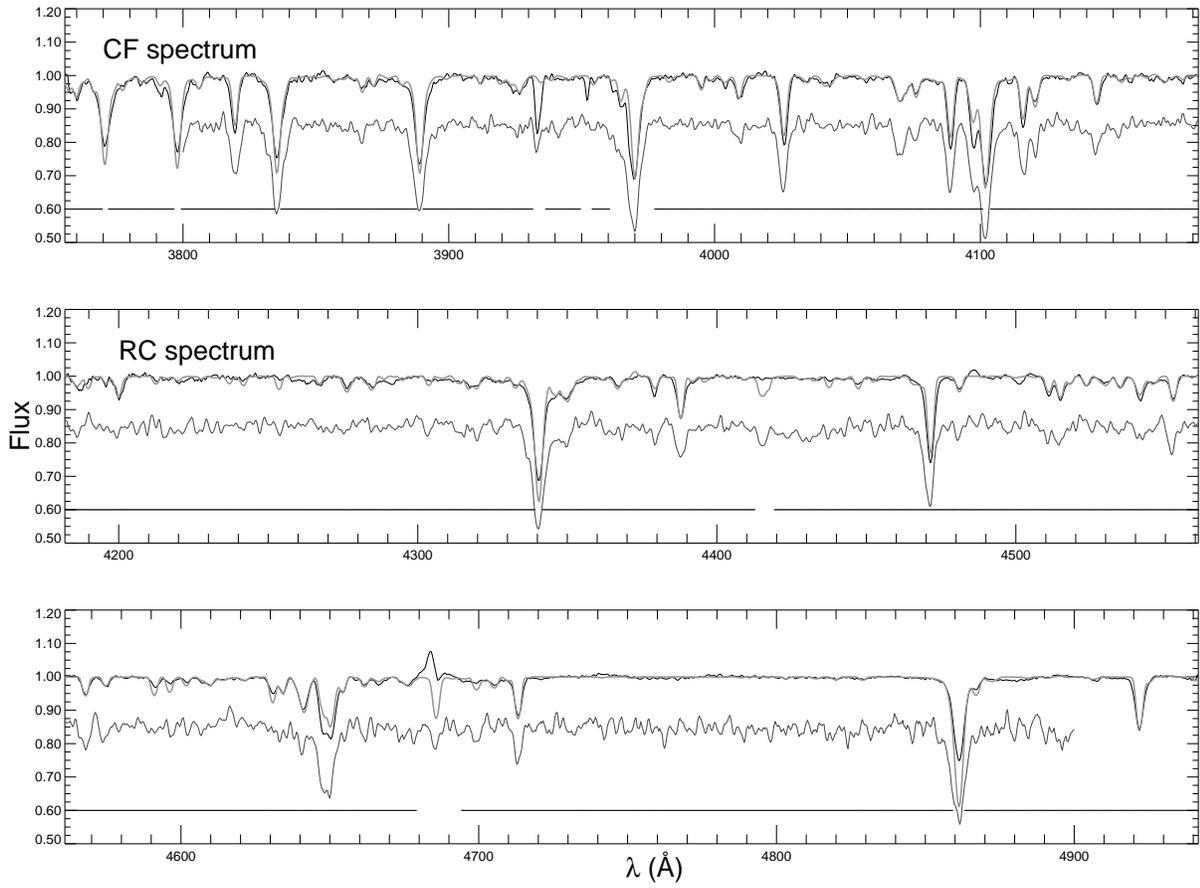}}
\end{center}
\caption{Optical spectra ({\it dark line}) together with a $T_{\rm eff} = 26$~kK and 
$\log g = 2.75$ TLUSTY CN model ({\it light line}) in the same format as Fig.~2.}
\label{fig3}
\end{figure}

\placefigure{fig3}     

The \ion{He}{1} and \ion{He}{2} line strengths are well matched in the
He enriched CN model spectra.  In particular the temperature sensitive
ratio of \ion{He}{2} $\lambda 4541$ to \ion{Si}{3} $\lambda 4552$
(equal for the O9.7 classification) is better reproduced by the
$T_{\rm eff} = 28$ kK and $\log g = 3.00$ model (Fig.~2) than the
$T_{\rm eff} = 26$ kK and $\log g = 2.75$ model (Fig.~3).  The
\ion{N}{3} $\lambda\lambda 4097,4379,4510,4514,4630,4634, 4640$ lines
are also reasonably well fit in the five times overabundant CN models.
On the other hand, O lines like \ion{O}{2} $\lambda\lambda
4069,4072,4075,4590,4596$ are too strong in the model spectra, which
suggests that the O abundance should be revised downwards from solar
values as expected for CNO-processed gas.  The other differences
between the observed and model spectra are related to the presence of
interstellar features (\ion{Ca}{2} $\lambda\lambda 3933, 3968$ in the
CF spectrum; most of the deep ISM features were removed from the RC
spectra).

In Figure~4 we present the averaged UV spectrum made with {\it
HST}/STIS with the best TLUSTY/SYNSPEC model superimposed as a lighter
line.  Figure~4 also includes an average UV spectrum of $\mu$~Nor,
based upon 34 high resolution, archival spectra from {\it IUE}.
Horizontal line segments indicate those regions where the lines
primarily originate in the photosphere, i.e., free from P~Cygni
stellar wind lines and from regions where interstellar lines were
removed by interpolation \citep{gie08}.  Overall, the line features in
the observed UV spectrum agree well with the model UV spectrum based
upon the optimal $T_{\rm eff}$ and $\log g$ parameters derived from
the optical and FUV spectral fits.  Note that the \ion{He}{2} $\lambda
1640$ feature appears in absorption as predicted, so there is no
evidence of the Raman scattering emission that was observed by
\citet{kap90} in the massive X-ray binary 4U1700--37.  There are,
however, a few specific regions where the match is less satisfactory.
For example, the blends surrounding \ion{Fe}{5} $\lambda$ 1422 and
\ion{Fe}{4} $\lambda\lambda 1596, 1615$ appear stronger in both the
spectra of HD~226868 and $\mu$~Nor, which suggests that the models are
underestimating the Fe line opacity in these wavelength regions.  The
\ion{S}{5} $\lambda 1502$ line \citep{how87} has a strength in the
spectrum of HD~226868 that falls between that of the model and of
$\mu$~Nor.  The deep feature near 1690 \AA~ is an instrumental flaw
near the edge of the detector at one grating tilt.

\begin{figure}
\begin{center}
{\includegraphics[angle=90,height=12cm]{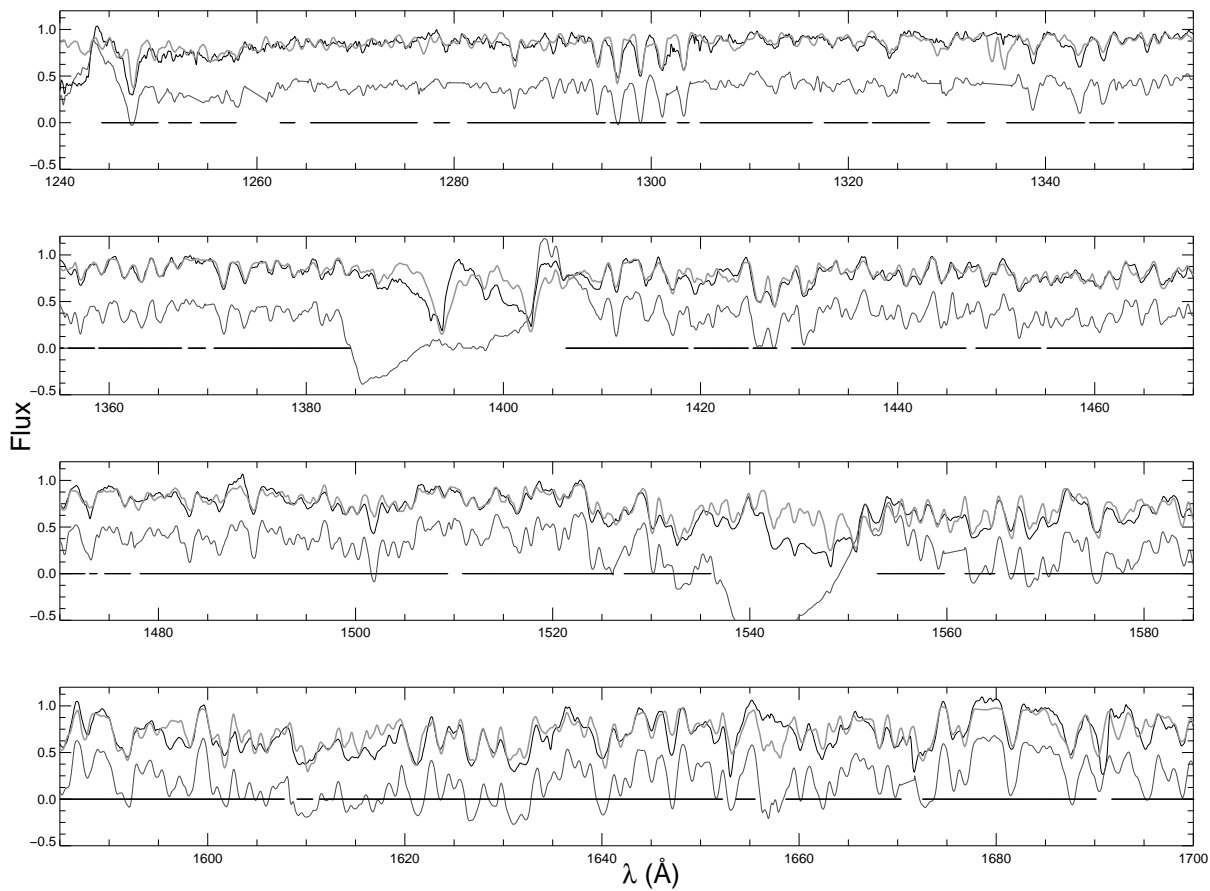}}
\end{center}
\caption{Observed UV spectrum ({\it dark line}) plus a TLUSTY CN model spectrum for
$T_{\rm eff} = 28$~kK and $\log g = 3.0$ ({\it light line}).  The spectrum of the
O9.7~Iab star $\mu$~Nor is offset by 0.6 from the model and HD~226868
spectra for comparison. The horizontal lines below the spectra indicate
regions included in the $\chi^2_\nu$ calculation.}
\label{fig4}
\end{figure}

\placefigure{fig4}     

There are huge variations in the stellar wind lines between the
orbital conjunctions that are due to X-ray ionization of the wind
\citep{gie08,vrt08}, and it is possible that X-ray heating might also
affect some of the photospheric lines.  Figure 5 compares the average
UV spectra at the two conjunction phases $\phi$ = 0.0 and 0.5
(inferior and superior conjunction of the supergiant, respectively).
With the exception of the known wind line changes, we find that the
spectra are almost identical between conjunctions.  Some slight
differences are seen in very strong features, such as the \ion{Si}{3}
$\lambda 1300$ complex and the \ion{Fe}{5} line blends in the 1600 --
1650 \AA ~region.  The deep lines appear somewhat deeper at $\phi =
0.0$ and have slightly extended blue wings compared to those observed
at $\phi = 0.5$ (when the black hole is in the foreground).  We
speculate that the deeper cores and blue extensions result from line
opacity that forms in the upper atmosphere where the outward wind
acceleration begins.  This outer part of the atmosphere in the
hemisphere facing the black hole may also experience X-ray ionization
(like the lower density wind) that promotes Si and Fe to higher
ionization levels and reduces the line opacity of the observed
transitions.

\begin{figure}
\begin{center}
{\includegraphics[angle=90,height=12cm]{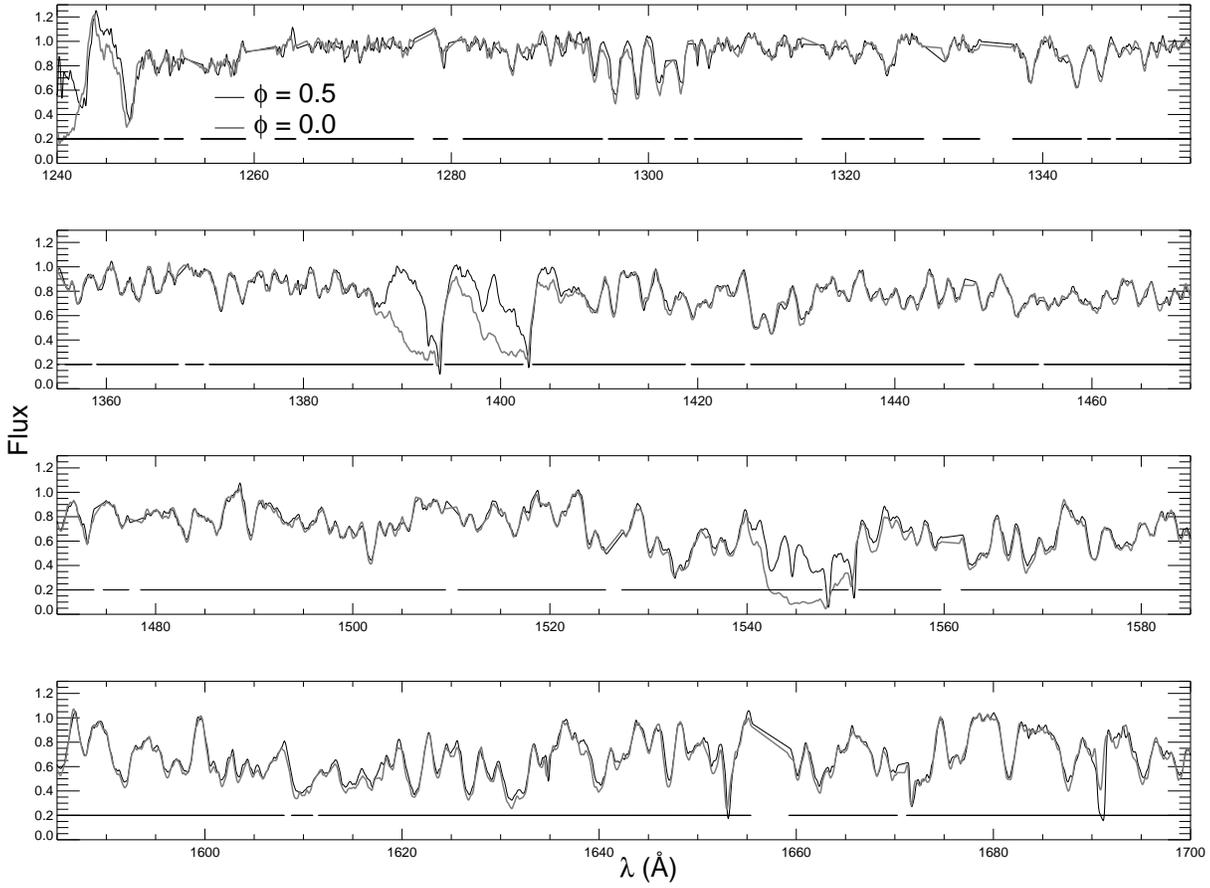}}
\end{center}
\caption{Averaged UV spectra of HD~226868 at orbital phases $\phi = 0.0$ ({\it light line})
and $\phi = 0.5$ ({\it dark line}). The horizontal line indicates those regions
without strong ISM features.}
\label{fig5}
\end{figure}

\placefigure{fig5}     

Our spectral fits are all based upon the existing OSTAR2002 and
BSTAR2006 grids, and it would certainly be worthwhile to explore more
specific models, for example, to derive reliable estimates of the He
and N overabundances.  A determination of the He abundance in
particular will be important for a definitive temperature estimate.
It is also important in such an analysis to consider the full effects
of the stellar wind in HD~226868.  \citet{her95} compared analyses of
the spectrum of HD~226868 from static, plane-parallel models with
unified, spherical models (that treat the photosphere and wind
together), and they found their $\log g$ estimate increased by about
0.2 dex (with no change in temperature) in the unified models.  Thus,
we suspect that our gravity estimate derived from the plane-parallel
TLUSTY code is probably a lower limit (approximately consistent with
the results of \citealt{her95} and \citealt{kar05}).


\section{UV -- IR Spectral Energy Distribution}     

We can use the derived model flux spectrum to fit the observed
spectral energy distribution (SED) and reassess the interstellar
extinction and the radius -- distance relation.  We collected the
archival low dispersion {\it IUE} spectra and combined these fluxes
with the {\it HST} spectra in wavelength bins spanning the FUV and NUV
regions.  We transformed the $UBV$ magnitudes from \citet{mas95} into
fluxes using the calibration of \citet{col96}, and the near-IR fluxes
were determined from a calibration of the 2MASS $JHK_s$ magnitudes
\citep{coh03,skr06}.  Then we fit the observed fluxes with the optimal
BSTAR2006 flux model (CN model, $\xi = 10$ km~s$^{-1}$, $T_{\rm eff} =
28$ kK, $\log g = 3.0$) to find the best reddening curve using the
extinction law from \citet{fit99}.  We placed additional weight on the
six optical and IR points to compensate for the larger number of UV
points.  Figure~6 shows the observed and best fit model fluxes for
HD~226868 that we obtained with a reddening $E(B-V) = 1.11\pm0.03$ mag
and a ratio of total to selective extinction $R_V = 3.02\pm 0.03$.
These values agree well with the previous reddening estimates that are
collected in Table~2.

\begin{figure}
\begin{center}
{\includegraphics[angle=90,height=12cm]{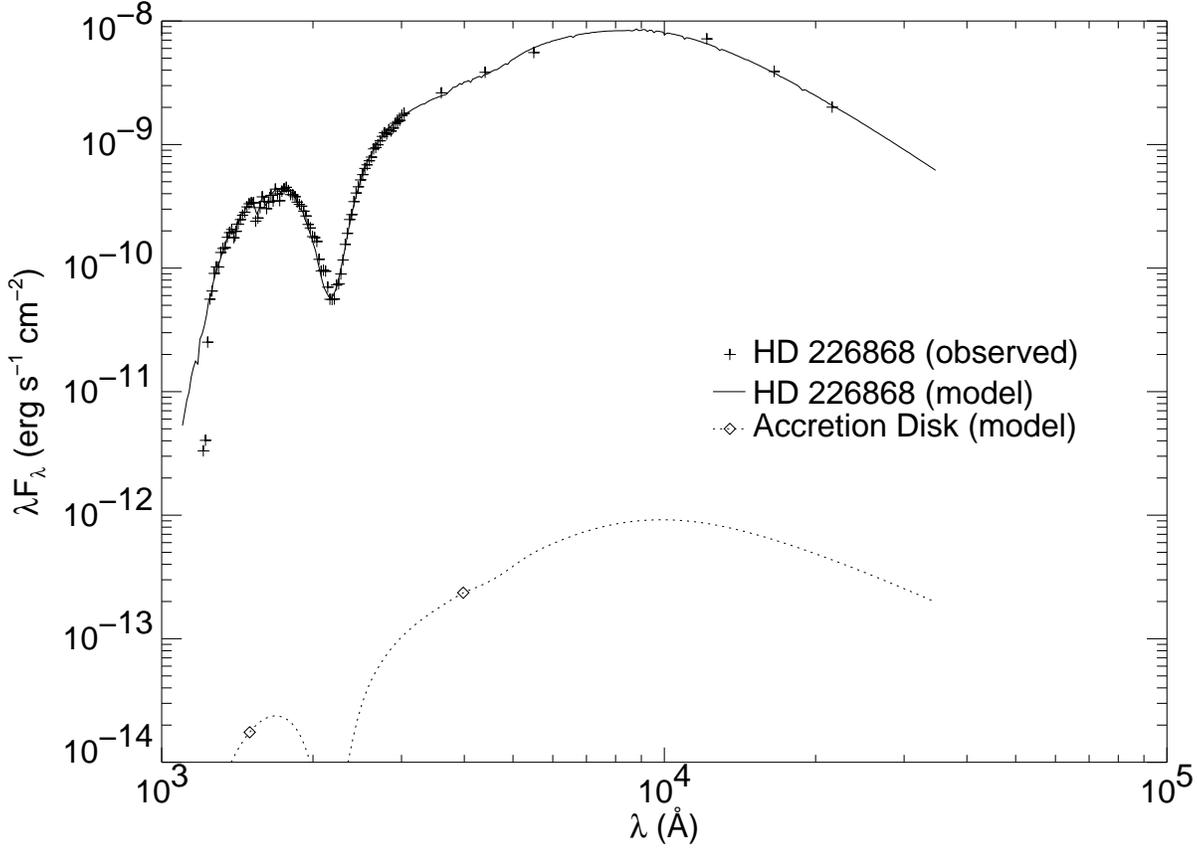}}
\end{center}
\caption{The spectral energy distribution of HD~226868 ({\it plus signs}) 
together and the TLUSTY best fit ({\it solid line}) for 
$T_{\rm eff} = 28$~kK and $\log g = 3.0$.  The UV points were binned from
the average {\it HST} and {\it IUE} spectra. The three optical points
are the $UBV$ measurements from \citet{mas95} and the three IR
points are taken from 2MASS.  Also shown is the extrapolation of the 
accretion disk flux model of \citet{mil02} ({\it dotted line}).}
\label{fig6}
\end{figure}

\placefigure{fig6}     

\placetable{tab2}      

For comparison we examined the colors and reddening of six field stars
within 10 arcminutes of HD~226868 in the sky.  These stars were
observed with the KPNO 4~m telescope and RC spectrograph using the
same blue region arrangement selected for our observations of
HD~226868 \citep{gie08}.  We made spectral classification of the
stars, and then we used the observed $UBV$ colors from \citet{mas95}
and the intrinsic color and absolute magnitude for the classification
\citep{gra92} to estimate reddening and distances to these stars.  Our
results are collected in Table~3 with the reddening estimate from
above listed for HD~226868.  \citet{bre73} estimated the distance of
HD~226868 as $d\approx 2.5$ kpc, and set a lower limit of 1~kpc based
upon the colors of other nearby field stars.  We find that there are
two stars at distances just under 1~kpc that have a similar reddening
to that of HD~226868, which is consistent with a distance to HD~226868
of $d\gtrsim 1.0$~kpc.

\placetable{tab3}      

The normalization of the fit to the SED yields the star's
limb-darkened angular diameter, $\theta = 96 \pm 6$ $\mu$as.  Then we
can calculate the luminosity and radius of the star as a function of
distance $d$ (in kpc) to HD~226868,
\begin{equation}\label{eq2}
 \frac{L_{1}}{L_{\odot}} = (5.9 \pm 2.1) \times 10^{4} d^{2},
\end{equation}
\begin{equation}\label{eq3}
 \frac{R_{1}}{R_{\odot}} = (10.3 \pm 0.7) d.
\end{equation}

It is important to check that these SED results are not affected by
long term or orbital flux variability, so we examined the archival
{\it IUE} low dispersion spectra \citep{gie08} and the {\it HST}
spectra to determine the amplitude of any flux variations in the UV.
We calculated the average continuum flux over three wavelength spans
(1252 -- 1380 \AA , 1410 -- 1350 \AA , and 1565 -- 1685 \AA) that
excluded the main wind features.  We then converted the UV fluxes to
differential magnitudes $\Delta m$.  We found no significant
differences between fluxes from times corresponding to the X-ray
low/hard state ({\it IUE}) and high/soft state ({\it IUE} and {\it
HST}; see \citealt{gie08} for X-ray state information), nor were there
any long term variations over the 25 year time span between the {\it
IUE} and {\it HST} observations.  On the other hand, we do find
marginal evidence of the orbital flux variations related to the tidal
distortion of the supergiant.  We plot in Figure~7 the mean orbital
flux variations of the three wavelength intervals for both {\it IUE}
and {\it HST} spectra that are averaged into eight bins of orbital
phase.  For comparison we also include the $V$-band ellipsoidal light
curve from \citet{kha81}.  The UV and $V$-band light curves appear to
have similar amplitudes, consistent with past estimates
\citep{tre80,vlo01}.  Note that the minima have approximately equal
depths (consistent with the optical results; \citealt{bal81}), which
suggests that there is little if any deep heating by X-rays of the
hemisphere of the supergiant facing the black hole.  Since the
amplitude of the light curve is small and the average UV fluxes
plotted in the SED in Figure~6 cover the full orbit, the ellipsoidal
variations have a minimal impact on the quantities derived from the
SED.

\begin{figure}
\begin{center}
{\includegraphics[angle=90,height=12cm]{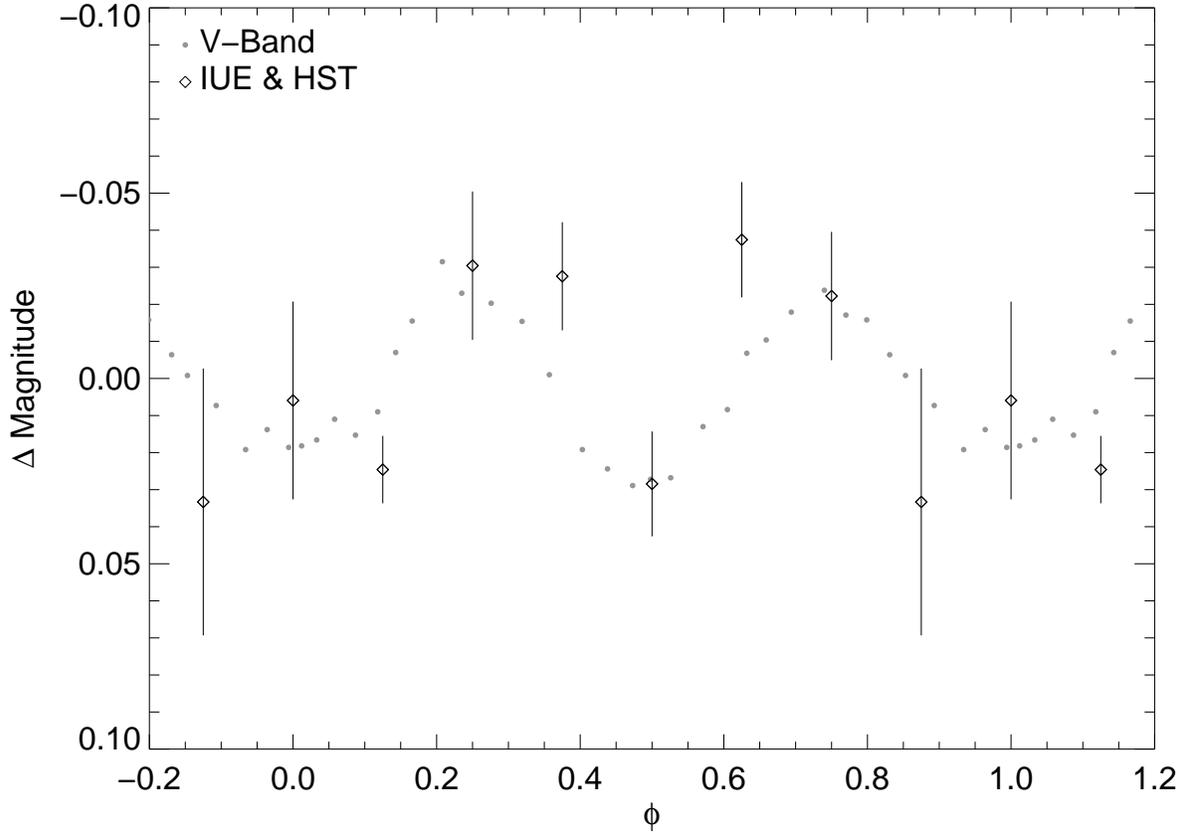}}
\end{center}
\caption{UV light curve from both the {\it HST} and {\it IUE} spectra
({\it diamonds}) compared with the $V$-band light curve ({\it circles}) 
from \citet{kha81}. The UV data were divided into eight orbital phase bins,
and the error bars indicate the standard deviation within each bin.}
\label{fig7}
\end{figure}

\placefigure{fig7}     

Finally, we need to consider if the SED has a non-stellar flux
contribution from the accretion disk around the black hole or from
other circumstellar gas.  \citet{bru78} estimated that the disk
contributes about 2\% of the optical flux, and there are reports of
small optical variations with superorbital periods that may correspond
to the precession of the accretion disk
\citep{kem87,bro99,szo07,pou08}.  Furthermore, \citet{dol01} observed
rapid UV variations that he argued originate in dying pulse trains of
infalling material passing the event horizon of the BH.  \citet{mil02}
developed a multi-color disk SED to model the X-ray continuum of
Cyg~X-1, and Dr.~Miller kindly sent us the model fluxes extrapolated
into the UV and optical.  These are also plotted in Figure~6 after
accounting for interstellar extinction.  Both the photospheric and
disk SEDs correspond to the Rayleigh-Jeans tail of a hot continuum,
and the model predicts that the disk contributes approximately 0.01\%
of the total flux in the UV to IR range.  This small fraction is
consistent with our successful fitting of the UV and optical line
features that would otherwise appear shallower by flux dilution if the
disk was a significant flux contributor.  Thus, our SED fitting is
probably unaffected by any non-stellar flux source.


\section{Mass of the Supergiant}     

In this section we will explore the mass consequences of our relations
for radius and luminosity as a function of distance.  \citet{pac74}
derived model-independent, minimum mass estimates for both components
as a function of distance based on the lack of X-ray eclipses (setting
a maximum orbital inclination) and the assumption that HD~226868 is
not larger than its Roche lobe (setting a lower limit on the ratio of
the supergiant to black hole mass, $M_1/M_2$).  We repeated his
analysis using our revised radius -- distance relationship (eq.\ [3]),
stellar effective temperature $T_{\rm eff} = 28$~kK, and current
values for the mass function $f(m)$ = 0.251$\pm$0.007 M$_{\odot}$ and
period $P$ = 5.599829 days \citep{gie03}.  The resulting minimum
masses are presented in columns 7 and 10 of Table~4 as a function of
distance $d$.

\placetable{tab4}      

  We can make further progress by assuming the supergiant has attained
synchronous rotation with the orbit since the stellar radius is
probably comparable in size to the Roche radius \citep{gie86b}.  We
take the ratio $\Omega$ of the star's spin angular velocity to orbital
angular velocity to be 1.  Then the projected rotational velocity
$V\sin i$ is related to the inclination $i$ by
\begin{equation}\label{eq4}
V\sin i = {{2 \pi}\over{P}} R_1 \sin i
\end{equation}
where $P$ is the orbital period.  The projected rotational velocity,
after correction for macroturbulent broadening, is estimated to be
$V\sin i = 95\pm 6$ km~s$^{-1}$ \citep{gie86a,nin87b,her95}. Inserting
equation~(\ref{eq3}) for $R_1$ we obtain an inclination estimate in terms of distance $d$ (kpc) of
\begin{equation}\label{eq5}
i = \arcsin ((1.02 \pm 0.09) d) .
\end{equation}
These inclination estimates are given in column 2 of Table~4.  Note
that this argument suggests a lower limit to the distance of $\approx
1.0$ kpc, similar to that found by reddening considerations.

\citet{gie86a} showed that the mass ratio can be estimated from the
ratio of the projected rotational velocity to the orbital
semiamplitude of the supergiant,
\begin{equation}\label{eq6}
{{V\sin i}\over{K}} = \rho (Q+1) \Phi(Q)
\end{equation}
where $\rho$ is the fill-out factor, i.e., the ratio of volume
equivalent radii of the star and Roche lobe, $Q=M_1/M_2$ is the mass
ratio, $\Phi$ is the ratio of the Roche lobe radius to the semimajor
axis \citep{egg83}, and synchronous rotation is assumed.  Thus, given
the observed values of $V\sin i$ and $K$ and an assumed value of
$\rho$, we can find the mass ratio and, with the inclination, the
masses of each star.  These masses are listed in columns 8, 9, 11 and
12 of Table~4 under headings that give the fill-out factor in
parentheses.  The run of masses is also shown in Figure~8 that
illustrates the mass solutions as a function of distance and fill-out
factor.  Loci of constant $\rho$ are denoted by dotted lines
(increasing right to left from $\rho=0.85$ to 1.0) while loci of
constant distance (and inclination angle) are shown by dashed lines.
The derived gravity values from these masses of $\log g \approx 3.3$
reinforces the idea that our spectral estimate of $\log g = 3.0$ is a
lower limit (see \S2).

We assumed synchronous rotation in the relations above because both
observations and theory indicate that the orbital synchronization time
scale in close binaries is shorter than the circularization time scale
\citep{cla95}, and since the orbit is circular, it follows that the
star must rotate at close to the synchronous rate.  However, it is
straight forward to see how the solutions will change if the
synchronism parameter $\Omega$ differs from unity.  In equation (5)
the distance $d$ can be replaced by the product $\Omega d$, while in
equation (6) the fill-out parameter $\rho$ can be replaced by
$\Omega\rho$.  If, for example, $\Omega = 0.95$, then the mass
solutions can be obtained from Table~4 and Figure~8 by selecting a
distance of $0.95 d$ and a fill-out ratio of $0.95 \rho$.

\placetable{tab4}      

\begin{figure}
\begin{center}
{\includegraphics[angle=90,height=12cm]{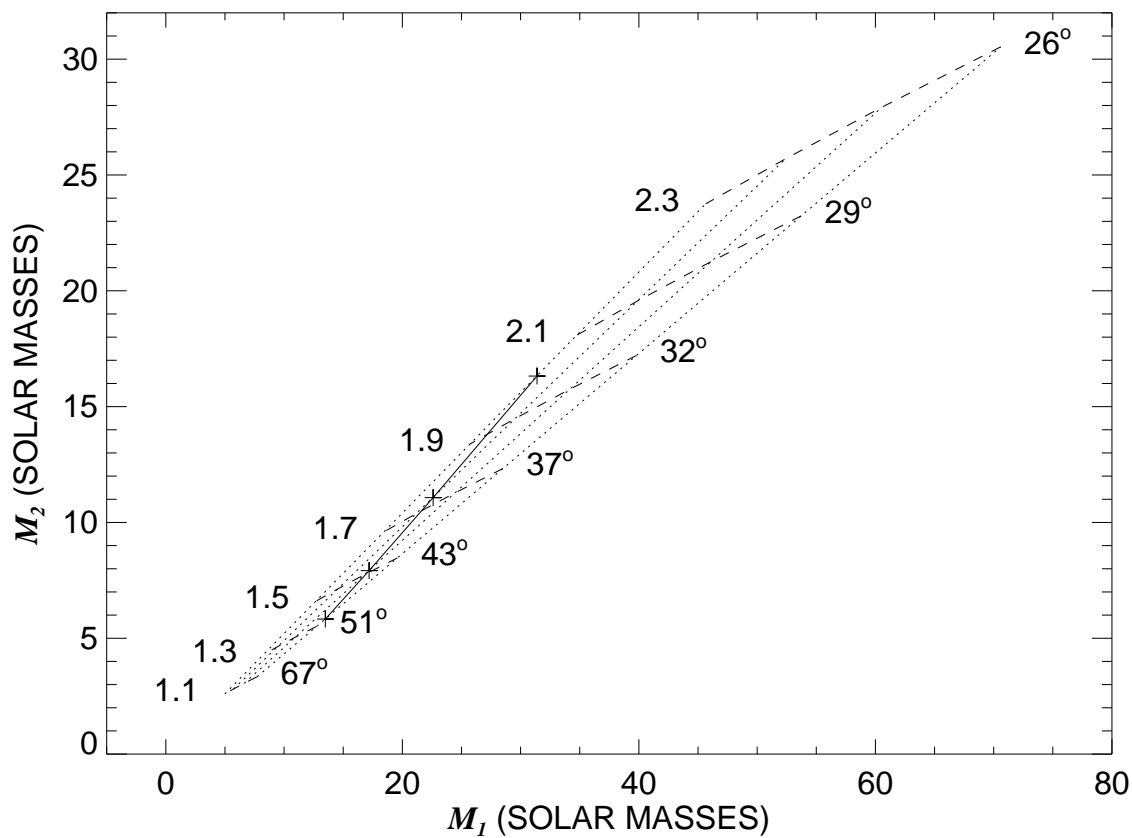}}
\end{center}
\caption{Mass plane diagram for the optical ($x$-axis) and X-ray ($y$-axis)
components assuming synchronous rotation of the supergiant. 
Dotted lines represent fill-out factors ($\rho$) of 0.85 to 1.0 in increments of 0.05 from
right-to-left, and the dashed lines show loci of constant distance (in kpc as labeled 
on the left side while the corresponding orbital inclination 
rounded to the nearest degree appears on the right side).
Plus signs connected by a solid line show the solutions that match the $V$-band
orbital light curve \citep{kha81} at each value of the fill-out factor.}
\label{fig8}
\end{figure}

\placefigure{fig8}     

The other important constraint comes from the ellipsoidal light curve.
The tidal distortion of the star results in a double-wave variation
(Fig.~7) whose amplitude depends on the inclination (maximal at
$i=90^\circ$) and degree of tidal distortion (maximal for fill-out
$\rho=1.0$).  In order to determine which parts of mass plane are
consistent with the observed variation, we constructed model $V$-band
light curves using the GENSYN code \citep{moc72,gie86a} for the four
values of fill-out factor illustrated in Figure~8.  There is a unique
solution for the best fit of the light curve along each line of
constant fill-out factor, since the light curve amplitude
monotonically decreases with decreasing inclination (increasing
distance).  The solid line in Figure~8 connects these best fit
solutions (indicated by plus sign symbols).  These light curve
solutions differ slightly from those presented by \citet{gie86a}
because we chose to fit the light curve from \citet{kha81} instead of
that from \citet{kem83}, and the differences in the solutions reflect
the uncertainties in the observed light curve.

There are several other constraints from hints about the mass transfer
process, luminosity, and distance that can provide additional limits
on the acceptable mass ranges.  Both \citet{gie86b} and \citet{nin87a}
presented arguments that the unusual \ion{He}{2} $\lambda 4686$
emission in the spectrum of HD~226868 originates in a tidal stream or
focused wind from the supergiant towards the black hole.  Furthermore,
\citet{gie86b} made radiative transfer calculations of the focused
wind emission profiles for models of the asymmetric wind from
\citet{fri82}, and they determined that the fill-out factor must
exceed $\rho=0.90$ in order to increase sufficiently the wind density
between the stars to account for the observed strength of the
\ion{He}{2} $\lambda 4686$ emission.  Thus, the presence of a focused
wind implies that the fill-out factor falls in the range $\rho=0.9 -
1.0$.

\citet{pac74} and \citet{zio05} argue that massive stars evolve at
near constant luminosity, and, therefore, the best solutions will obey
the observed mass -- luminosity relation.  Table~4 lists the derived
luminosity $\log L_{1}$ (column 4) as a function of distance (eq.\
[2]) plus the predicted luminosities for the mass solutions determined
for the $\rho=0.9$ and 1.0 cases, $\log L_{1}^\star(0.9)$ and $\log
L_{1}^\star(1.0)$, respectively (columns 5 and 6).  These predictions
are based upon the mass -- luminosity relations for $T_{\rm
eff}=28$~kK stars from the model evolutionary sequences made by
\citet{sch92}.  We find that the observed and predicted luminosities
match over the distance range of $d=1.7$ ($\rho=0.9$) to 2.0 kpc
($\rho=1.0$), closer than the range advocated by \citet{zio05} who
adopted a higher temperature and hence higher luminosity.  Note that
some stars in mass transfer binaries appear overluminous for their
mass, so these distances should probably be considered as upper
limits.

Several authors have suggested that the position and proper motion of
HD~226868 indicates that it is a member of the Cyg~OB3 association
\citep{mir03} that has a distance of 1.6-2.5 kpc \citep{uya01}.
However, a radio parallax study by \citet{les99} indicates a smaller
(but possibly consistent) distance of $1.4^{+0.9}_{-0.4}$~kpc for Cyg
X-1.  Our fits of the ellipsoidal light curve suggest that the maximum
allowable distance is $d\approx 2.0$~kpc (for $\rho=1.0$) The
interstellar reddening indicates a distance of at least 1.0~kpc (\S3),
which is probably consistent with the strength of interstellar
\ion{Ca}{2} lines.  \citet{meg05} present a method for determining the
distance to O supergiants using the equivalent width $W_{\lambda}$ of
the \ion{Ca}{2} $\lambda 3933$ feature.  Using their calibration with
the value of $W_{\lambda} = 400 \pm 10$ m\AA ~from \citet{gie86a}
yields a distance $d = 1.2$~kpc.  Since the reddening of HD~226868 is
approximately the same as that for the much more distant Cepheid,
V547~Cyg \citep{bre73}, the ISM must have a relatively low density
beyond $\approx 1$~kpc along this line of sight through the Galaxy, so
we suspect that the distance derived from the interstellar \ion{Ca}{2}
line is probably a lower limit.

All of these constraints are consistent with the mass solutions for a
fill-out factor range of $\rho = 0.9 - 1.0$, and the corresponding
mass ranges are listed in Table~5.  We also list mass estimates from
earlier investigations.  Our downward revision of the effective
temperature results in lower luminosity estimates than adopted by
\citet{zio05}, and consequently, our mass estimates (based upon the
light curve) are significantly lower than his mass estimates (based
upon the mass -- luminosity relation from models).  In fact, the lower
limit for the black hole mass now overlaps comfortably with the mass
determined by \citet{sha07} using the correlation between the X-ray
quasi-periodic oscillation frequency and spectral index, so the
apparent discrepancy in black hole mass estimates from X-ray and
optical data is now resolved.  If the X-ray derived mass is accurate,
then the mass solution for fill-out factor $\rho=0.91$ is preferred
($M_1 = 19 M_\odot$ and the distance is $d=1.6$~kpc).

\placetable{tab5}      

Our analysis of the first high resolution UV spectra of HD~226868 and
of the complementary optical spectra shows that the photospheric line
spectrum can be matched by adopting an atmosphere mixed with
CNO-processed gas with an effective temperature $T_{\rm eff} = 28.0
\pm 2.5$~kK and log $g \gtrsim 3.0 \pm 0.25$.  Assuming synchronous
rotation ($\Omega = 1$) and using the fill-out factor range from
above, the mass of the supergiant ranges from $M_1 = 17 - 31
M_{\odot}$ and the black hole mass ranges from $M_2 = 8 -
16M_{\odot}$. This corresponds to an inclination of $i = 31^\circ -
43^\circ$ and a distance of $d = 1.5 - 2.0$ kpc.  Better estimates of
the masses may be possible in the future.  For example, both the {\it
GAIA} \citep{jor08} and {\it SIM Lite} \citep{unw08} space astrometry
missions will provide an accurate parallax and distance.  Furthermore,
pointed observations with SIM Lite will measure the astrometric motion
of the supergiant around the system center of mass, yielding
independent estimates of both the orbital inclination and distance (by
equating the astrometric and radial velocity semi-major axes;
\citealt{tom09}).  Finally, future high dispersion X-ray spectroscopy
with the {\it International X-ray
Observatory}\footnote{http://ixo.gsfc.nasa.gov/index.html} will
measure the orbital motion of the black hole through the orbital
Doppler shifts of accretion disk flux in the Fe K$\alpha$ line
\citep{mil07}.  By comparing the optical and X-ray orbital velocity
curves, we will have a secure mass ratio that, together with the
distance estimate, will lead to unique and accurate mass
determinations of the supergiant and black hole.


\acknowledgments

We thank the staffs of the Kitt Peak National Observatory and the
Space Telescope Science Institute (STScI) for their support in
obtaining these observations.  We are also grateful to Thierry Lanz
and Ivan Hubeny for information about their model atmosphere grid and
to Jon Miller for providing us with details about his accretion disk
model.  Support for {\it HST} proposal number GO-9840 was provided by
NASA through a grant from the Space Telescope Science Institute, which
is operated by the Association of Universities for Research in
Astronomy, Incorporated, under NASA contract NAS5-26555.  The {\it
IUE} data presented in this paper were obtained from the Multimission
Archive at the Space Telescope Science Institute (MAST).  Support for
MAST for non-HST data is provided by the NASA Office of Space Science
via grant NAG5-7584 and by other grants and contracts.  This
publication makes use of data products from the Two Micron All Sky
Survey, which is a joint project of the University of Massachusetts
and the Infrared Processing and Analysis Center/California Institute
of Technology, funded by the National Aeronautics and Space
Administration and the National Science Foundation.  Bolton's research
is partially supported by a Natural Sciences and Engineering Research
Council of Canada (NSERC) Discovery Grant.  Hadrava's research is
funded under grant projects GA\v{C}R 202/06/0041 and LC06014.  Herrero
thanks the Spanish MEC for support under project AY 2007-67456-C02-01.
This work was also supported by the National Science Foundation under
grants AST-0205297, AST-0506573, and AST-0606861.  Institutional
support has been provided from the GSU College of Arts and Sciences
and from the Research Program Enhancement fund of the Board of Regents
of the University System of Georgia, administered through the GSU
Office of the Vice President for Research.  We are grateful for all
this support.



\bibliographystyle{apj}
\bibliography{apj-jour,paper}

\begin{thebibliography}{65}
\expandafter\ifx\csname natexlab\endcsname\relax\def\natexlab#1{#1}\fi

\bibitem[{{Abubekerov} {et~al.}(2005){Abubekerov}, {Antokhina}, \&
  {Cherepashchuk}}]{abu05}
{Abubekerov}, M.~K., {Antokhina}, {\'E}.~A., \& {Cherepashchuk}, A.~M. 2005,
  Astronomy Reports, 49, 801

\bibitem[{{Balog} {et~al.}(1981){Balog}, {Goncharskij}, \&
  {Cherepashchuk}}]{bal81}
{Balog}, N.~I., {Goncharskij}, A.~V., \& {Cherepashchuk}, A.~M. 1981, Soviet
  Astronomy Letters, 7, 336

\bibitem[{{Bolton}(1972)}]{bol72}
{Bolton}, C.~T. 1972, \nat, 240, 124

\bibitem[{{Bregman} {et~al.}(1973){Bregman}, {Butler}, {Kemper}, {Koski},
  {Kraft}, \& {Stone}}]{bre73}
{Bregman}, J., {Butler}, D., {Kemper}, E., {Koski}, A., {Kraft}, R.~P., \&
  {Stone}, R.~P.~P. 1973, \apjl, 185, L117

\bibitem[{{Brocksopp} {et~al.}(1999){Brocksopp}, {Fender}, {Larionov}, {Lyuty},
  {Tarasov}, {Pooley}, {Paciesas}, \& {Roche}}]{bro99}
{Brocksopp}, C., {Fender}, R.~P., {Larionov}, V., {Lyuty}, V.~M., {Tarasov},
  A.~E., {Pooley}, G.~G., {Paciesas}, W.~S., \& {Roche}, P. 1999, \mnras, 309,
  1063

\bibitem[{{Bruevich} {et~al.}(1978){Bruevich}, {Kiliachkov}, {Syunyaev}, \&
  {Shevchenko}}]{bru78}
{Bruevich}, V.~V., {Kiliachkov}, N.~N., {Syunyaev}, R.~A., \& {Shevchenko},
  V.~S. 1978, Soviet Astronomy Letters, 4, 292

\bibitem[{{Canalizo} {et~al.}(1995){Canalizo}, {Koenigsberger}, {Pe{\~n}a}, \&
  {Ruiz}}]{can95}
{Canalizo}, G., {Koenigsberger}, G., {Pe{\~n}a}, D., \& {Ruiz}, E. 1995,
  Revista Mexicana de Astronom{\'i}a y Astrof{\'i}sica, 31, 63

\bibitem[{{Claret} {et~al.}(1995){Claret}, {Gim{\'e}nez}, \& {Cunha}}]{cla95}
{Claret}, A., {Gim{\'e}nez}, A., \& {Cunha}, N.~C.~S. 1995, \aap, 299, 724

\bibitem[{{Cohen} {et~al.}(2003){Cohen}, {Wheaton}, \& {Megeath}}]{coh03}
{Cohen}, M., {Wheaton}, W.~A., \& {Megeath}, S.~T. 2003, \aj, 126, 1090

\bibitem[{{Colina} {et~al.}(1996){Colina}, {Bohlin}, \& {Castelli}}]{col96}
{Colina}, L., {Bohlin}, R., \& {Castelli}, F. 1996, HST Instrument Science
  Report CAL/SCS-008 (Baltimore: STScI)

\bibitem[{{Dolan}(2001)}]{dol01}
{Dolan}, J.~F. 2001, \pasp, 113, 974

\bibitem[{{Eggleton}(1983)}]{egg83}
{Eggleton}, P.~P. 1983, \apj, 268, 368

\bibitem[{{Fitzpatrick}(1999)}]{fit99}
{Fitzpatrick}, E.~L. 1999, \pasp, 111, 63

\bibitem[{{Friend} \& {Castor}(1982)}]{fri82}
{Friend}, D.~B., \& {Castor}, J.~I. 1982, \apj, 261, 293

\bibitem[{{Gies} \& {Bolton}(1986{\natexlab{a}})}]{gie86a}
{Gies}, D.~R., \& {Bolton}, C.~T. 1986{\natexlab{a}}, \apj, 304, 371

\bibitem[{{Gies} \& {Bolton}(1986{\natexlab{b}})}]{gie86b}
------. 1986{\natexlab{b}}, \apj, 304, 389

\bibitem[{{Gies} {et~al.}(2008){Gies}, {Bolton}, {Blake}, {Caballero-Nieves},
  {Crenshaw}, {Hadrava}, {Herrero}, {Hillwig}, {Howell}, {Huang}, {Kaper},
  {Koubsk{\'y}}, \& {McSwain}}]{gie08}
{Gies}, D.~R. {et~al.} 2008, \apj, 678, 1237

\bibitem[{{Gies} {et~al.}(2003){Gies}, {Bolton}, {Thomson}, {Huang}, {McSwain},
  {Riddle}, {Wang}, {Wiita}, {Wingert}, {Cs{\'a}k}, \& {Kiss}}]{gie03}
------. 2003, \apj, 583, 424

\bibitem[{{Gray}(1992)}]{gra92}
{Gray}, D.~F. 1992, {The Observation and Analysis of Stellar Photospheres}, 2nd
  edn. ({Cambridge}: Cambridge Univ. Press)

\bibitem[{{Herrero} {et~al.}(1995){Herrero}, {Kudritzki}, {Gabler}, {Vilchez},
  \& {Gabler}}]{her95}
{Herrero}, A., {Kudritzki}, R.~P., {Gabler}, R., {Vilchez}, J.~M., \& {Gabler},
  A. 1995, \aap, 297, 556

\bibitem[{{Howarth}(1987)}]{how87}
{Howarth}, I.~D. 1987, \mnras, 226, 249

\bibitem[{{Jordan}(2008)}]{jor08}
{Jordan}, S. 2008, Astronomische Nachrichten, 329, 875

\bibitem[{{Kaper} {et~al.}(1990){Kaper}, {Hammerschlag-Hensberge}, \&
  {Takens}}]{kap90}
{Kaper}, L., {Hammerschlag-Hensberge}, G., \& {Takens}, R.~J. 1990, \nat, 347,
  652

\bibitem[{{Karitskaya} {et~al.}(2005){Karitskaya}, {Agafonov}, {Bochkarev},
  {Bondar}, {Galazutdinov}, {Lee}, {Musaev}, {Sapar}, {Sharova}, \&
  {Shimanskii}}]{kar05}
{Karitskaya}, E.~A. {et~al.} 2005, Astronomical and Astrophysical Transactions,
  24, 383

\bibitem[{{Karitskaya} {et~al.}(2008){Karitskaya}, {Bochkarev}, {Bondar'},
  {Galazutdinov}, {Lee}, {Musaev}, {Sapar}, \& {Shimanskii}}]{kar08}
{Karitskaya}, E.~A., {Bochkarev}, N.~G., {Bondar'}, A.~V., {Galazutdinov},
  G.~A., {Lee}, B.-C., {Musaev}, F.~A., {Sapar}, A.~A., \& {Shimanskii}, V.~V.
  2008, Astronomy Reports, 52, 362

\bibitem[{{Kemp} {et~al.}(1983){Kemp}, {Barbour}, {Henson}, {Kraus}, {Nolt},
  {Radostitz}, {Priedhorsky}, {Terrell}, \& {Walker}}]{kem83}
{Kemp}, J.~C. {et~al.} 1983, \apjl, 271, L65

\bibitem[{{Kemp} {et~al.}(1987){Kemp}, {Karitskaya}, {Kumsiashvili}, {Lyutyi},
  {Khruzina}, \& {Cherepashchuk}}]{kem87}
{Kemp}, J.~C., {Karitskaya}, E.~A., {Kumsiashvili}, M.~I., {Lyutyi}, V.~M.,
  {Khruzina}, T.~S., \& {Cherepashchuk}, A.~M. 1987, Soviet Astronomy, 31, 170

\bibitem[{{Khaliullin} \& {Khaliullina}(1981)}]{kha81}
{Khaliullin}, K.~F., \& {Khaliullina}, A.~I. 1981, Soviet Astronomy, 25, 593

\bibitem[{{Lanz} \& {Hubeny}(2003)}]{lan03}
{Lanz}, T., \& {Hubeny}, I. 2003, \apjs, 146, 417

\bibitem[{{Lanz} \& {Hubeny}(2007)}]{lan07}
------. 2007, \apjs, 169, 83

\bibitem[{{Lefever}(2007)}]{lef07}
{Lefever}, K. 2007, Ph.D. Thesis (Katholieke Universiteit Leuven)

\bibitem[{{Lestrade} {et~al.}(1999){Lestrade}, {Preston}, {Jones}, {Phillips},
  {Rogers}, {Titus}, {Rioja}, \& {Gabuzda}}]{les99}
{Lestrade}, J.-F., {Preston}, R.~A., {Jones}, D.~L., {Phillips}, R.~B.,
  {Rogers}, A.~E.~E., {Titus}, M.~A., {Rioja}, M.~J., \& {Gabuzda}, D.~C. 1999,
  \aap, 344, 1014

\bibitem[{{Martins} {et~al.}(2005){Martins}, {Schaerer}, \& {Hillier}}]{mar05}
{Martins}, F., {Schaerer}, D., \& {Hillier}, D.~J. 2005, \aap, 436, 1049

\bibitem[{{Massey} {et~al.}(1995){Massey}, {Johnson}, \&
  {Degioia-Eastwood}}]{mas95}
{Massey}, P., {Johnson}, K.~E., \& {Degioia-Eastwood}, K. 1995, \apj, 454, 151

\bibitem[{{Megier} {et~al.}(2005){Megier}, {Strobel}, {Bondar}, {Musaev},
  {Han}, {Kre{\l}owski}, \& {Galazutdinov}}]{meg05}
{Megier}, A., {Strobel}, A., {Bondar}, A., {Musaev}, F.~A., {Han}, I.,
  {Kre{\l}owski}, J., \& {Galazutdinov}, G.~A. 2005, \apj, 634, 451

\bibitem[{{Miller}(2007)}]{mil07}
{Miller}, J.~M. 2007, \araa, 45, 441

\bibitem[{{Miller} {et~al.}(2002){Miller}, {Fabian}, {Wijnands}, {Remillard},
  {Wojdowski}, {Schulz}, {Di Matteo}, {Marshall}, {Canizares}, {Pooley}, \&
  {Lewin}}]{mil02}
{Miller}, J.~M. {et~al.} 2002, \apj, 578, 348

\bibitem[{{Mirabel} \& {Rodrigues}(2003)}]{mir03}
{Mirabel}, I.~F., \& {Rodrigues}, I. 2003, Science, 300, 1119

\bibitem[{{Mochnacki} \& {Doughty}(1972)}]{moc72}
{Mochnacki}, S.~W., \& {Doughty}, N.~A. 1972, \mnras, 156, 51

\bibitem[{{Ninkov} {et~al.}(1987{\natexlab{a}}){Ninkov}, {Walker}, \&
  {Yang}}]{nin87a}
{Ninkov}, Z., {Walker}, G.~A.~H., \& {Yang}, S. 1987{\natexlab{a}}, \apj, 321,
  438

\bibitem[{{Ninkov} {et~al.}(1987{\natexlab{b}}){Ninkov}, {Walker}, \&
  {Yang}}]{nin87b}
------. 1987{\natexlab{b}}, \apj, 321, 425

\bibitem[{{Osterbrock} \& {Ferland}(2006)}]{ost06}
{Osterbrock}, D.~E., \& {Ferland}, G.~J. 2006, {Astrophysics of Gaseous Nebulae
  and Active Galactic Nuclei}, 2nd edn. ({Sausalito, CA}: {University Science
  Books})

\bibitem[{{Paczy\'{n}ski}(1974)}]{pac74}
{Paczy\'{n}ski}, B. 1974, \aap, 34, 161

\bibitem[{{Poutanen} {et~al.}(2008){Poutanen}, {Zdziarski}, \&
  {Ibragimov}}]{pou08}
{Poutanen}, J., {Zdziarski}, A.~A., \& {Ibragimov}, A. 2008, \mnras, 389, 1427

\bibitem[{{Repolust} {et~al.}(2004){Repolust}, {Puls}, \& {Herrero}}]{rep04}
{Repolust}, T., {Puls}, J., \& {Herrero}, A. 2004, \aap, 415, 349

\bibitem[{{Ryans} {et~al.}(2002){Ryans}, {Dufton}, {Rolleston}, {Lennon},
  {Keenan}, {Smoker}, \& {Lambert}}]{rya02}
{Ryans}, R.~S.~I., {Dufton}, P.~L., {Rolleston}, W.~R.~J., {Lennon}, D.~J.,
  {Keenan}, F.~P., {Smoker}, J.~V., \& {Lambert}, D.~L. 2002, \mnras, 336, 577

\bibitem[{{Savage} {et~al.}(1985){Savage}, {Massa}, {Meade}, \&
  {Wesselius}}]{sav85}
{Savage}, B.~D., {Massa}, D., {Meade}, M., \& {Wesselius}, P.~R. 1985, \apjs,
  59, 397

\bibitem[{{Schaller} {et~al.}(1992){Schaller}, {Schaerer}, {Meynet}, \&
  {Maeder}}]{sch92}
{Schaller}, G., {Schaerer}, D., {Meynet}, G., \& {Maeder}, A. 1992, \aaps, 96,
  269

\bibitem[{{Searle} {et~al.}(2008){Searle}, {Prinja}, {Massa}, \&
  {Ryans}}]{sea08}
{Searle}, S.~C., {Prinja}, R.~K., {Massa}, D., \& {Ryans}, R. 2008, \aap, 481,
  777

\bibitem[{{Shaposhnikov} \& {Titarchuk}(2007)}]{sha07}
{Shaposhnikov}, N., \& {Titarchuk}, L. 2007, \apj, 663, 445

\bibitem[{{Skrutskie} {et~al.}(2006){Skrutskie}, {Cutri}, {Stiening},
  {Weinberg}, {Schneider}, {Carpenter}, {Beichman}, {Capps}, {Chester},
  {Elias}, {Huchra}, {Liebert}, {Lonsdale}, {Monet}, {Price}, {Seitzer},
  {Jarrett}, {Kirkpatrick}, {Gizis}, {Howard}, {Evans}, {Fowler}, {Fullmer},
  {Hurt}, {Light}, {Kopan}, {Marsh}, {McCallon}, {Tam}, {Van Dyk}, \&
  {Wheelock}}]{skr06}
{Skrutskie}, M.~F. {et~al.} 2006, \aj, 131, 1163

\bibitem[{{Szostek} \& {Zdziarski}(2007)}]{szo07}
{Szostek}, A., \& {Zdziarski}, A.~A. 2007, \mnras, 375, 793

\bibitem[{{Tomsick} {et~al.}(2009){Tomsick}, {Shaklan}, \& {Pan}}]{tom09}
{Tomsick}, J., {Shaklan}, S., \& {Pan}, X. 2009, {in SIMLite Astrometric
  Observatory}, ed. J.~{Davidson}, S.~{Edberg}, R.~{Danner}, B.~{Nemati}, \&
  S.~{Unwin} ({Pasadena: NASA JPL 400-1360}), 97

\bibitem[{{Treves} {et~al.}(1980){Treves}, {Chiappetti}, {Tanzi}, {Tarenghi},
  {Gursky}, {Dupree}, {Hartmann}, {Raymond}, {Davis}, \& {Black}}]{tre80}
{Treves}, A. {et~al.} 1980, \apj, 242, 1114

\bibitem[{{Unwin} {et~al.}(2008){Unwin}, {Shao}, {Tanner}, {Allen}, {Beichman},
  {Boboltz}, {Catanzarite}, {Chaboyer}, {Ciardi}, {Edberg}, {Fey}, {Fischer},
  {Gelino}, {Gould}, {Grillmair}, {Henry}, {Johnston}, {Johnston}, {Jones},
  {Kulkarni}, {Law}, {Majewski}, {Makarov}, {Marcy}, {Meier}, {Olling}, {Pan},
  {Patterson}, {Pitesky}, {Quirrenbach}, {Shaklan}, {Shaya}, {Strigari},
  {Tomsick}, {Wehrle}, \& {Worthey}}]{unw08}
{Unwin}, S.~C. {et~al.} 2008, \pasp, 120, 38

\bibitem[{{Uyaniker} {et~al.}(2001){Uyaniker}, {F{\"u}rst}, {Reich},
  {Aschenbach}, \& {Wielebinski}}]{uya01}
{Uyaniker}, B., {F{\"u}rst}, E., {Reich}, W., {Aschenbach}, B., \&
  {Wielebinski}, R. 2001, \aap, 371, 675

\bibitem[{{van Loon} {et~al.}(2001){van Loon}, {Kaper}, \&
  {Hammerschlag-Hensberge}}]{vlo01}
{van Loon}, J.~T., {Kaper}, L., \& {Hammerschlag-Hensberge}, G. 2001, \aap,
  375, 498

\bibitem[{{Vrtilek} {et~al.}(2008){Vrtilek}, {Boroson}, {Hunacek}, {Gies}, \&
  {Bolton}}]{vrt08}
{Vrtilek}, S.~D., {Boroson}, B.~S., {Hunacek}, A., {Gies}, D., \& {Bolton},
  C.~T. 2008, \apj, 678, 1248

\bibitem[{{Wade} \& {Rucinski}(1985)}]{wad85}
{Wade}, R.~A., \& {Rucinski}, S.~M. 1985, \aaps, 60, 471

\bibitem[{{Walborn}(1973)}]{wal73}
{Walborn}, N.~R. 1973, \apjl, 179, L123

\bibitem[{{Walborn} \& {Fitzpatrick}(1990)}]{wal90}
{Walborn}, N.~R., \& {Fitzpatrick}, E.~L. 1990, \pasp, 102, 379

\bibitem[{{Webster} \& {Murdin}(1972)}]{web72}
{Webster}, B.~L., \& {Murdin}, P. 1972, \nat, 235, 37

\bibitem[{{Wegner}(2002)}]{weg02}
{Wegner}, W. 2002, Baltic Astronomy, 11, 1

\bibitem[{{Wu} {et~al.}(1982){Wu}, {Eaton}, {Holm}, {Milgrom}, \&
  {Hammerschlag-Hensberge}}]{wu82}
{Wu}, C.-C., {Eaton}, J.~A., {Holm}, A.~V., {Milgrom}, M., \&
  {Hammerschlag-Hensberge}, G. 1982, \pasp, 94, 149

\bibitem[{{Zi{\'o}{\l}kowski}(2005)}]{zio05}
{Zi{\'o}{\l}kowski}, J. 2005, \mnras, 358, 851

\end{thebibliography}


\clearpage

\begin{deluxetable}{ccccccccccccc}
\tablewidth{0pc}
\tabletypesize{\scriptsize}
\rotate
\tablenum{1}
\tablecaption{$\chi^2_\nu$ for Spectral Fits with Models\label{tab1}}
\tablewidth{0pt}
\tablehead{
\colhead{Spectrum} & 
\colhead{$\log g$} & 
\colhead{Model}    & 
\multicolumn{10}{c}{$T_{\rm eff}$ (kK)} \\
\noalign{\smallskip}
\cline{4-13}
\noalign{\smallskip}
\colhead{Source} & 
\colhead{(cm s$^{-2}$)} & 
\colhead{Code} & 
\colhead{\phn24.0} & 
\colhead{\phn25.0} & 
\colhead{\phn26.0} & 
\colhead{\phn27.0} &
\colhead{\phn27.5} &  
\colhead{\phn28.0} &
\colhead{\phn29.0} &
\colhead{\phn30.0} &  
\colhead{\phn32.5} &  
\colhead{\phn35.0}}  
\startdata
HST & 4.00 & O   &\nodata &\nodata &\nodata &\nodata &\phn35.6&\nodata &\nodata &\phn24.4&\phn24.1&\phn29.3\\
HST & 3.75 & O   &\nodata &\nodata &\nodata &\nodata &\phn30.7&\nodata &\nodata &\phn23.0&\phn24.7&\phn33.4\\
HST & 3.50 & O   &\nodata &\nodata &\nodata &\nodata &\phn26.0&\nodata &\nodata &\phn22.6&\phn27.0&\phn42.8\\
HST & 3.25 & O   &\nodata &\nodata &\nodata &\nodata &\phn22.8&\nodata &\nodata &\phn23.8&\phn36.2&\phn77.5\\
HST & 3.00 & O   &\nodata &\nodata &\nodata &\nodata &\phn23.0&\nodata &\nodata &\phn32.8&\nodata &\nodata \\
HST & 3.00 & B   &\phn39.5&\phn29.1&\phn24.1&\phn23.0&\nodata &\phn24.1&\phn27.3&\phn34.8&\nodata &\nodata \\
HST & 3.00 & BCN &\phn42.5&\phn31.7&\phn25.5&\phn23.6&\nodata &\phn23.9&\phn25.9&\phn31.6&\nodata &\nodata \\
HST & 2.75 & B   &\phn28.5&\phn25.9&\phn27.0&\phn31.7&\nodata &\nodata &\nodata &\nodata &\nodata &\nodata \\
HST & 2.75 & BCN &\phn31.6&\phn26.5&\phn26.2&\phn28.5&\nodata &\nodata &\nodata &\nodata &\nodata &\nodata \\
CF  & 4.00 & O   &\nodata &\nodata &\nodata &\nodata &   114.5&\nodata &\nodata &\phn72.2&\phn53.0&\phn43.7\\
CF  & 3.75 & O   &\nodata &\nodata &\nodata &\nodata &\phn87.8&\nodata &\nodata &\phn55.8&\phn36.7&\phn43.5\\
CF  & 3.50 & O   &\nodata &\nodata &\nodata &\nodata &\phn59.3&\nodata &\nodata &\phn29.6&\phn30.7&\phn53.9\\
CF  & 3.25 & O   &\nodata &\nodata &\nodata &\nodata &\phn32.5&\nodata &\nodata &\phn21.2&\phn47.7&\phn85.5\\
CF  & 3.00 & O   &\nodata &\nodata &\nodata &\nodata &\phn19.6&\nodata &\nodata &\phn51.0&\nodata &\nodata \\
CF  & 3.00 & B   &\phn51.0&\phn36.9&\phn24.4&\phn16.7&\nodata &\phn16.2&\phn27.9&\phn50.0&\nodata &\nodata \\
CF  & 3.00 & BCN &\phn59.2&\phn44.8&\phn30.6&\phn19.4&\nodata &\phn12.9&\phn13.5&\phn31.4&\nodata &\nodata \\
CF  & 2.75 & B   &\phn22.4&\phn17.3&\phn23.7&\phn51.9&\nodata &\nodata &\nodata &\nodata &\nodata &\nodata \\
CF  & 2.75 & BCN &\phn28.7&\phn17.2&\phn12.5&\phn20.2&\nodata &\nodata &\nodata &\nodata &\nodata &\nodata \\
RC  & 4.00 & O   &\nodata &\nodata &\nodata &\nodata &\phn85.9&\nodata &\nodata &\phn42.0&\phn25.1&\phn42.7\\
RC  & 3.75 & O   &\nodata &\nodata &\nodata &\nodata &\phn68.3&\nodata &\nodata &\phn31.9&\phn22.7&\phn61.3\\
RC  & 3.50 & O   &\nodata &\nodata &\nodata &\nodata &\phn50.0&\nodata &\nodata &\phn20.8&\phn38.9&\phn92.5\\
RC  & 3.25 & O   &\nodata &\nodata &\nodata &\nodata &\phn33.6&\nodata &\nodata &\phn24.5&\phn85.5&   143.4\\
RC  & 3.00 & O   &\nodata &\nodata &\nodata &\nodata &\phn25.2&\nodata &\nodata &\phn90.5&\nodata &\nodata \\
RC  & 3.00 & B   &\phn88.1&\phn61.3&\phn35.9&\phn19.8&\nodata &\phn19.6&\phn51.5&   106.6&\nodata &\nodata \\
RC  & 3.00 & BCN &   112.9&\phn83.8&\phn53.7&\phn29.7&\nodata &\phn15.6&\phn17.9&\phn62.6&\nodata &\nodata \\
RC  & 2.75 & B   &\phn47.7&\phn25.1&\phn29.4&\phn95.7&\nodata &\nodata &\nodata &\nodata &\nodata &\nodata \\
RC  & 2.75 & BCN &\phn69.2&\phn36.3&\phn17.1&\phn26.3&\nodata &\nodata &\nodata &\nodata &\nodata &\nodata \\
\enddata
\end{deluxetable}

\begin{deluxetable}{lll}
\tablewidth{0pc}
\tabletypesize{\scriptsize}
\tablenum{2}
\tablecaption{Interstellar Reddening Estimates\label{tab2}}
\tablewidth{0pt}
\tablehead{
\colhead{Source} & 
\colhead{$E(B-V)$} & 
\colhead{$R_{V}$}}
\startdata
\citet{bre73} & 1.12 (5)   &  3.0     \\
\citet{tre80} & 1.06       &  3.0     \\
\citet{wu82}  & 0.95 (7)   &  3.1     \\
\citet{sav85} & 1.080 (25) &  3.1     \\
\citet{weg02} & 1.03       &  3.32    \\
This Paper    & 1.11 (3)   &  3.02 (3)\\
\enddata
\tablecomments{Numbers in parentheses give the error in the last digit quoted.}
\end{deluxetable}

\begin{deluxetable}{ccccccc}
\tablewidth{0pc}
\tabletypesize{\scriptsize}
\tablenum{3}
\tablecaption{Interstellar Reddening for HD~226868 and Nearby Stars \label{tab3}}
\tablewidth{0pt}
\tablehead{
\colhead{R.A. (2000)} & 
\colhead{Dec. (2000)} & 
\colhead{Spectral} & 
\colhead{$m_{V}$} & 
\colhead{$m_{B}$} & 
\colhead{$E(B-V)$} &  
\colhead{$d$} \\
\colhead{(hh mm ss.ss)} &
\colhead{(dd mm ss.s)} &
\colhead{Classification} &
\colhead{(mag)} &
\colhead{(mag)} &
\colhead{(mag)} &
\colhead{(kpc)}}
\startdata 
19 58 21.68 & +35 12 05.8 &  O9.7 Iab &\phn8.81 &\phn9.64 &  1.11  & \nodata \\
19 58 23.44 & +35 14 32.2 &    A3 V   &  15.15  &  16.46  &  1.26  &  0.98 \\
19 58 29.31 & +35 09 27.5 &    G0 V   &  15.11  &  16.35  &  0.66  &  0.54 \\
19 58 04.44 & +35 11 48.3 &    G2 V   &  15.06  &  15.97  &  0.28  &  0.82 \\
19 58 02.06 & +35 14 00.0 &    F2 V   &  15.37  &  16.74  &  1.02  &  0.73 \\
19 58 57.44 & +35 05 31.7 &    G2 V   &  15.42  &  16.43  &  0.38  &  0.84 \\
19 58 44.45 & +35 08 09.8 &    F5 V   &  15.76  &  16.87  &  0.68  &  1.07 \\
\enddata
\end{deluxetable}

\begin{deluxetable}{cccccccccccc}
\tablewidth{0pc}
\tabletypesize{\scriptsize}
\rotate
\tablenum{4}
\tablecaption{Mass and Luminosity versus Distance for HD~226868 \label{tab4}}
\tablewidth{0pt}
\tablehead{
\colhead{$d$} & 
\colhead{$i$} & 
\colhead{$R_{1}$} & 
\colhead{log $L_{1}$} & 
\colhead{log $L_{1}^{\star}$(0.9)}  & 
\colhead{log $L_{1}^{\star}$(1.0)}  & 
\colhead{$M_{1}^{min}$} &
\colhead{$M_{1}^{sync}$(0.9)} &  
\colhead{$M_{1}^{sync}$(1.0)} &  
\colhead{$M_{2}^{min}$} &
\colhead{$M_{2}^{sync}$(0.9)} &
\colhead{$M_{2}^{sync}$(1.0)} \\
\colhead{(kpc)} & 
\colhead{($^{\circ}$)} & 
\colhead{($R_{\odot}$)} & 
\colhead{($L_{\odot}$)} & 
\colhead{($L_{\odot}$)} & 
\colhead{($L_{\odot}$)} & 
\colhead{($M_{\odot}$)} &
\colhead{($M_{\odot}$)} &  
\colhead{($M_{\odot}$)} &
\colhead{($M_{\odot}$)} &
\colhead{($M_{\odot}$)} &  
\colhead{($M_{\odot}$)}}  
\startdata
1.1 & 67.5 & 11.4 & 4.85 & 3.06 & 2.77 &\phn5.0&\phn6.6&\phn5.0&\phn2.7&\phn3.1&\phn2.6\\
1.2 & 57.9 & 12.4 & 4.93 & 3.41 & 3.04 &\phn6.4&\phn8.6&\phn6.5&\phn3.1&\phn4.0&\phn3.4\\
1.3 & 51.4 & 13.4 & 5.00 & 3.83 & 3.35 &\phn7.9&  10.9 &\phn8.2&\phn3.5&\phn5.0&\phn4.3\\
1.4 & 46.5 & 14.5 & 5.06 & 4.31 & 3.71 &\phn9.6&  13.7 &  10.3 &\phn4.0&\phn6.3&\phn5.4\\
1.5 & 42.6 & 15.5 & 5.12 & 4.70 & 4.13 &  11.6 &  16.8 &  12.7 &\phn4.5&\phn7.7&\phn6.6\\
1.6 & 39.4 & 16.5 & 5.18 & 4.99 & 4.57 &  13.8 &  20.4 &  15.4 &\phn5.0&\phn9.4&\phn8.0\\
1.7 & 36.7 & 17.6 & 5.23 & 5.21 & 4.85 &  16.3 &  24.5 &  18.4 &\phn5.5&  11.3 &\phn9.6\\
1.8 & 34.4 & 18.6 & 5.28 & 5.37 & 5.08 &  19.0 &  29.0 &  21.9 &\phn6.1&  13.4 &  11.4 \\
1.9 & 32.3 & 19.6 & 5.33 & 5.54 & 5.26 &  21.9 &  34.1 &  25.7 &\phn6.6&  15.7 &  13.4 \\
2.0 & 30.5 & 20.7 & 5.37 & 5.68 & 5.40 &  25.2 &  39.8 &  30.0 &\phn7.2&  18.4 &  15.6 \\
2.1 & 28.9 & 21.7 & 5.41 & 5.80 & 5.56 &  28.7 &  46.1 &  34.7 &\phn7.9&  21.3 &  18.1 \\
2.2 & 27.5 & 22.7 & 5.45 & 5.92 & 5.69 &  32.5 &  53.0 &  39.9 &\phn8.5&  24.4 &  20.8 \\
2.3 & 26.2 & 23.8 & 5.49 & 6.05 & 5.79 &  36.7 &  60.6 &  45.6 &\phn9.2&  27.9 &  23.8 \\
2.4 & 25.0 & 24.8 & 5.53 & 6.17 & 5.90 &  41.1 &  68.8 &  51.9 &\phn9.9&  31.7 &  27.0 \\
2.5 & 24.0 & 25.9 & 5.57 & 6.27 & 6.02 &  45.9 &  77.8 &  58.6 &  10.6 &  35.9 &  30.5 \\
\enddata
\end{deluxetable}

\begin{deluxetable}{lcc}
\tablewidth{0pc}
\tabletypesize{\scriptsize}
\tablenum{5}
\tablecaption{Mass Estimates \label{tab5}}
\tablehead{
\colhead{} &
\colhead{$M_{1}$} &
\colhead{$M_{2}$} \\
\colhead{Source} &
\colhead{($M_{\odot}$)} &
\colhead{($M_{\odot}$)}}
\startdata
\citet{bal81}  &  20 -- 27     &   7 -- 12    \\
\citet{gie86a} &  23 -- 38     &  10 -- 20    \\
\citet{nin87b} &  20           &  10          \\
\citet{her95}  &  17.8         &  10.1        \\
\citet{abu05}  &  22           &  8.2 -- 12.8 \\
\citet{zio05}  &  30 -- 50     &  13.5 -- 29  \\
\citet{sha07}  &  \nodata      &  7.9 -- 9.5  \\
This paper     &  17 -- 31     &  8 -- 16     \\
\enddata
\end{deluxetable}

\clearpage



\input{epsf}

\clearpage

\clearpage
 
\clearpage
 
\clearpage
 
\clearpage
 
\clearpage
 
\clearpage
 
\end{document}